\documentclass[final,5p,times]{elsarticle}

\usepackage{graphics}
\usepackage{float}
\usepackage{epstopdf}
\usepackage{color}
\usepackage{subfigure}
\usepackage{graphicx}
\usepackage{amssymb}
\usepackage{eqnarray}
\usepackage{lipsum}
\usepackage{amsmath,amsfonts,amsthm,bm} 

\usepackage[amssymb,thinqspace,thinspace]{SIunits}

\journal{journal}

\begin{document}

	\begin{frontmatter}
	\title{Warren-Averbach line broadening analysis from a time of flight neutron diffractometer}
\author[label1,label2]{D.M. Collins}
\cortext[label3]{d.m.collins@bham.ac.uk}

\author[label2]{A.J. Wilkinson}
\author[label2]{R.I. Todd}
\cortext[label4]{richard.todd@materials.ox.ac.uk}

\address[label1]{School of Metallurgy \& Materials, University of Birmingham, Birmingham, B15 2TT, UK}
\address[label2]{Department of Materials, University of Oxford, Parks Road, Oxford, OX1 3PH, UK}

	\begin{abstract}
The well known Warren-Averbach theory of diffraction line profile broadening is shown to be applicable to time of flight data obtained from a neutron spallation source. Without modification, the method is applied to two very different examples; a cold worked ferritic steel and a thermally stressed alumina-30\% SiC composite. Values of root mean square strains averaged over a range of lengths for the ferritic steel were used to estimate dislocation densities; values were found to be in good agreement with geometrically necessary dislocation densities independently measured from similarly orientated grains measured from electron backscatter diffraction analysis. An analytical model for the ceramic is described to validate the estimate of root mean square strain. 

	\end{abstract}


	\end{frontmatter}

\section{Introduction}

The seminal theory by Warren and Averbach \cite{Warren1950,Warren1952,WarrenProgMetal} to describe the plastic deformation of metals provides a method to quantify crystallite size and strain from broadened diffraction line profiles. Size broadening can arise from both coherent and incoherently diffracting domains, as discovered earlier by Scherrer \cite{Scherrer1918}. Such domains are microstructural features that may comprise grains, twins, stacking faults, second phase particles, subgrain or deformation mosaicity structures. Strain broadening results from imperfections in the diffracting lattice that, themselves, generate inhomogenous crystalline distortions. These typically arise from point or line defects. As the broadening nature is different for size and strain, the simultaneous effects can be quantified independently. The method is hence a powerful characterisation tool that yields a compendium of information from bulk specimens that has the generality to be valid for various material classes including ceramics, polymers, composites as well as metals.

Separation of size and strain broadening components has long been known to be obtainable from X-ray diffraction line profile analysis, but has not been extended to time-of-flight neutron diffraction. This paper adopts the original Warren and Averbach line broadening methodology to demonstrate that suitably high resolution time of flight data are also well suited to such analysis. A derivation of the powder theory for time of flight diffraction is given, followed by example applications of the Warren-Averbach methodology to time of flight data obtained from (1) a cold worked steel in which line broadening is caused by dislocation generation, and (2) an alumina-30\% SiC ceramic with spatially varying thermal residual microstresses resulting from the thermal expansion mismatch between the phases during cooling after sintering. To demonstrate the validity of the results, comparisons are made to geometrically necessary dislocation (GND) densities obtained from electron backscatter diffraction (EBSD) analysis for the steel and an analytical model for the ceramic.

\section{Theory}

We follow the Warren-Averbach analysis \cite{Warren1950, WarrenProgMetal} adapting it where appropriate for use with a neutron time-of-flight diffractometer. The analysis considers the $00l$ reflection of an orthorhombic material \cite{Stokes1944a}. This simplifies the analysis, but it can be shown that the results apply to any set of planes in any crystal structure \cite{Stokes1944b, Warren1955}. A single distorted crystal is considered in which the position ${\bf{R}}_{m_1 m_2 m_3}$ of each unit cell is specified in terms of the (undistorted) orthorhombic lattice vectors $\bf{a}_1$, $\bf{a}_2$, $\bf{a}_3$, plus a displacement ${\bm{\delta}}_{m_1 m_2 m_3}$ from the ideal position in the undistorted crystal:
\begin{equation}
\label{RIT1}
{\bf{R}}_{m_1 m_2 m_3} = m_1 {\bf{a}}_1 + m_2 {\bf{a}}_2 + m_3 {\bf{a}}_3 + {\bm{\delta}}_{m_1 m_2 m_3}
\end{equation}

\noindent where $m_1$, $m_2$ and $m_3$ are integers specifying a particular cell. 

If the diffraction vector is represented in terms of the reciprocal lattice vectors ${\bf{b}}_1$, ${\bf{b}}_2$, ${\bf{b}}_3$:
\begin{equation}
\label{RIT2}
\frac{{\bf{s}}-{\bf{s}}_0}{\lambda} = h_1 {\bf{b}}_1 + h_2 {\bf{b}}_2 + h_3 {\bf{b}}_3
\end{equation}

\noindent where $h_1$, $h_2$ and $h_3$ are continuous variables and $\bf{s}$ and $\bf{s}_0$ are unit vectors parallel to the incident and diffracted beams respectively, then the total scattered intensity $I$ relative to that from the reference scattering length is:
\begin{equation}
\begin{split}
\label{RIT3}
I(h_1 h_2 h_3) & = F_S^2 \sum_{m_1} \sum_{m_2} \sum_{m_3} \sum_{m_1'} \sum_{m_2'} \sum_{m_3'} \exp \bigg[ 2\pi i \bigg\{ h_1(m_1 - m_1')  \\
& + h_2(m_2 - m_2') + h_3(m_3 - m_3') + \left( \frac{\bf{s}- \bf{s}_0}{\lambda} \right) \\
& \cdot ({\bm{\delta}}_{m_1 m_2 m_3}-{\bm{\delta}}_{m_1' m_2' m_3'}) \bigg\} \bigg]
\end{split}
\end{equation}

\noindent where $F_S$ is the structure factor.

A single small element of a neutron detector array in a time-of-flight diffractometer measures the intensity of scattered neutrons with a fixed $2\theta$  and azimuthal angle, and varying $\lambda$. The corresponding diffraction vector therefore has varying length along a fixed direction. The interference function given in Eq. \ref{RIT3} represents the spreading of the reciprocal lattice point of an individual crystal in reciprocal space. To obtain the measured intensity, it is necessary to sum the contributions from all the crystals in the sample. It is assumed in what follows that the extent of the spreading is small compared with the length of the diffraction vector and that the effects of any crystallographic texture or other source of anisotropy are insignificant over the small range of orientations concerned.


Figure \ref{RIT_Fig} shows a scattering vector of length $h_3\bf{b}_3$ whose direction has minimum distance $r$ from the centre of a $00l$ reciprocal lattice point a crystal. The number of crystals with $h_3\bf{b}_3$ passing through an annulus centred on the scattering vector direction, with thickness d$r$ and radius $r$ is:
\begin{equation}
\label{RIT4}
M\frac{2\pi r \rm{d} r}{4\pi h_3^2 b_3^2} = M\frac{r \rm{d} r}{2 h_3^2 b_3^2}
\end{equation}

\noindent where $M$ is the product of the total number of crystals and the multiplicity of the reflection. The average intensity contributed by each of the crystals with all possible rotations $\omega$ around a line joining the lattice point to the origin is:
\begin{equation}
\label{RIT5}
\frac{1}{2\pi} \int_0^{2\pi} I(r,\omega,h_3) \,\rm{d}\omega~~.
\end{equation}

\noindent The total intensity sampled at the tip of the diffraction vector is therefore obtained by summing over $r$ to include all crystals:
\begin{equation}
\begin{split}
\label{RIT6}
I(h_3) & = \iint \frac{I(rwh_3)Mr \, \rm{d}\omega \, \rm{d}r}{4\pi h_3^2 b_3^2} \\
           & = \iint \frac{I(h_1h_2h_3)M b_1 b_2 {\rm{d}}h_1 {\rm{d}} h_2 }{4\pi h_3^2 b_3^2} \\
           & = \iint \frac{I(h_1h_2h_3)M {\rm{d}}h_1 {\rm{d}} h_2 } {4\pi v_{\rm{a}} h_3^2 b_3^3}
\end{split}
\end{equation}

\noindent where $v_{\rm a}$ is the volume of the unit cell.

Using the de Broglie relation, the length of the reciprocal lattice vector for a neutron can be written:
\begin{equation}
\label{RIT7}
h_3b_3 = \frac{2\sin\theta}{\lambda} = \frac{2 m L \sin \theta}{h_{\rm p} t}
\end{equation}

\noindent where m is the neutron mass, $h_{\rm p}$, is Planck's constant, $L$ is the neutron path length and $t$ is the time-of-flight. Substituting for $I(h_1 h_2 h_3)$ and $h_3$ in Eq. \ref{RIT6} using Eqs. \ref{RIT3} and \ref{RIT7} gives:
\begin{equation}
\begin{split}
\label{RIT8}
I(h_3) & = \frac{M h_{\rm p}^2 t^2 F_S^2 }{16\pi v_{\rm a} m (L \sin \theta)^2 b_3} \times \frac{1}{F_S^2} \iint I(h_1h_2h_3)\,{\rm{d}}h_1 \, {\rm{d}} h_2 \\ 
& = K \iint \sum_{m_1} \sum_{m_2} \sum_{m_3} \sum_{m_1'} \sum_{m_2'} \sum_{m_3'} \exp \bigg[ 2\pi i \bigg\{ h_1(m_1 - m1')  \\
& ~~~~~~~~~~+ h_2(m_2 - m_2') + h_3(m_3 - m_3') + \left( \frac{\bf{s}- \bf{s}_0}{\lambda} \right) \\
& \cdot ({\bm{\delta}}_{m_1 m_2 m_3}-{\bm{\delta}}_{m_1' m_2' m_3'}) \bigg\} \bigg] {\rm d} h_1 {\rm d}h_2~~.
\end{split}
\end{equation}

\noindent where $K = \frac{M h_{\rm p}^2 t^2 F_S^2 }{16\pi v_{\rm a} m (L \sin \theta)^2 b_3} $. The important result of this analysis is that  $K$ is approximately constant for a time of flight diffractometer. This is because $t$ and $F_S$ vary little close to the reflection and the measured intensities are normalised to account for the variation in incident intensity with wavelength. For a single elemental neutron detector, $L\sin\theta$ is obviously constant, but in practice, extended detector banks covering a range of $2\theta$ and $L$ are used to increase the total neutron count. These detectors are ``focused'', however, either geometrically or electronically, such that $L\sin\theta$ remains constant, and Eq. \ref{RIT7} shows that $t$ is also constant for a given diffraction vector (or $d$-spacing). The Warren-Averbach analysis can therefore be used without significant modification for extended detector banks providing that anisotropic effects in the specimen remain small over the additional range of lattice plane orientations sampled.

Eq. \ref{RIT8} gives the intensity measured by a perfect (i.e. free from instrumental broadening) neutron time-of-flight diffractometer as a function of $h_3$, which can be calculated easily for a particular diffraction peak from the time-of-flight or equivalent $d$-spacing data provided at neutron sources as:
\begin{equation}
\label{RIT9}
h_3 = \frac{t_0 l}{t} = \frac{d_0 l}{d}
\end{equation}

\noindent where $d$ is the interplanar spacing which would give a Bragg peak at a particular $t$, and $t_0$ and $d_0$ are the values of $t$ and $d$ respectively for the centroids of the peak being analysed. In reality, instrumental broadening is significant and needs to be removed from the results as described in Section 3.2.

With the exception of $K$, the right hand side of Eq. \ref{RIT8} concerns only the distorted crystal and is identical to the right hand side of Eq. \ref{RIT5} in \cite{Warren1950}. Warren \cite{WarrenProgMetal} goes on to show that, with reasonable approximations, this can be written as a Fourier series:
\begin{equation}
\label{RIT10}
I(h_3) = KN \sum_{n = - \infty} ^{\infty} \big\{A_n \cos 2\pi n h_3 + B_n \sin 2\pi n h_3 \big\}
\end{equation}

\noindent where $N$ is the total number of unit cells per crystal. The cosine and sine coefficients $A_n$ and $B_n$ are given by:
\begin{equation}
\label{RIT11}
A_n= \frac{N_n}{N_3} \langle \cos 2 \pi l Z_n \rangle
\end{equation}
\begin{equation}
\label{RIT12}
B_n= \frac{N_n}{N_3} \langle \sin 2 \pi l Z_n \rangle
\end{equation}

\noindent in which $N_3$ is the average number of unit cells per crystal in a column normal to the diffracting planes, $N_n$ is the average number of cells possessing an $n^{\rm th}$ nearest neighbour cell in the same column, and the displacement along the column length between $n^{th}$ nearest neighbours due to the distortion is $Z_na_3$. Only the cosine coefficients $A_n$ are used in the Warren-Averbach analysis; the information obtainable from the sine coefficients $B_n$ is less useful and in any case the broadening is usually close to being symmetrical so that analysis around the peak centre gives $B_n$ values close to zero.

The Fourier coefficients of the physically broadening profile are a product of size, $A_n^S$, and distortion, $A_n^D$ coefficients, giving 
\begin{equation}
\label{FourierProduct}
A_n= A_n^S A_n^D
\end{equation}


\begin{figure*}
   \centering
  \includegraphics[width=\textwidth]{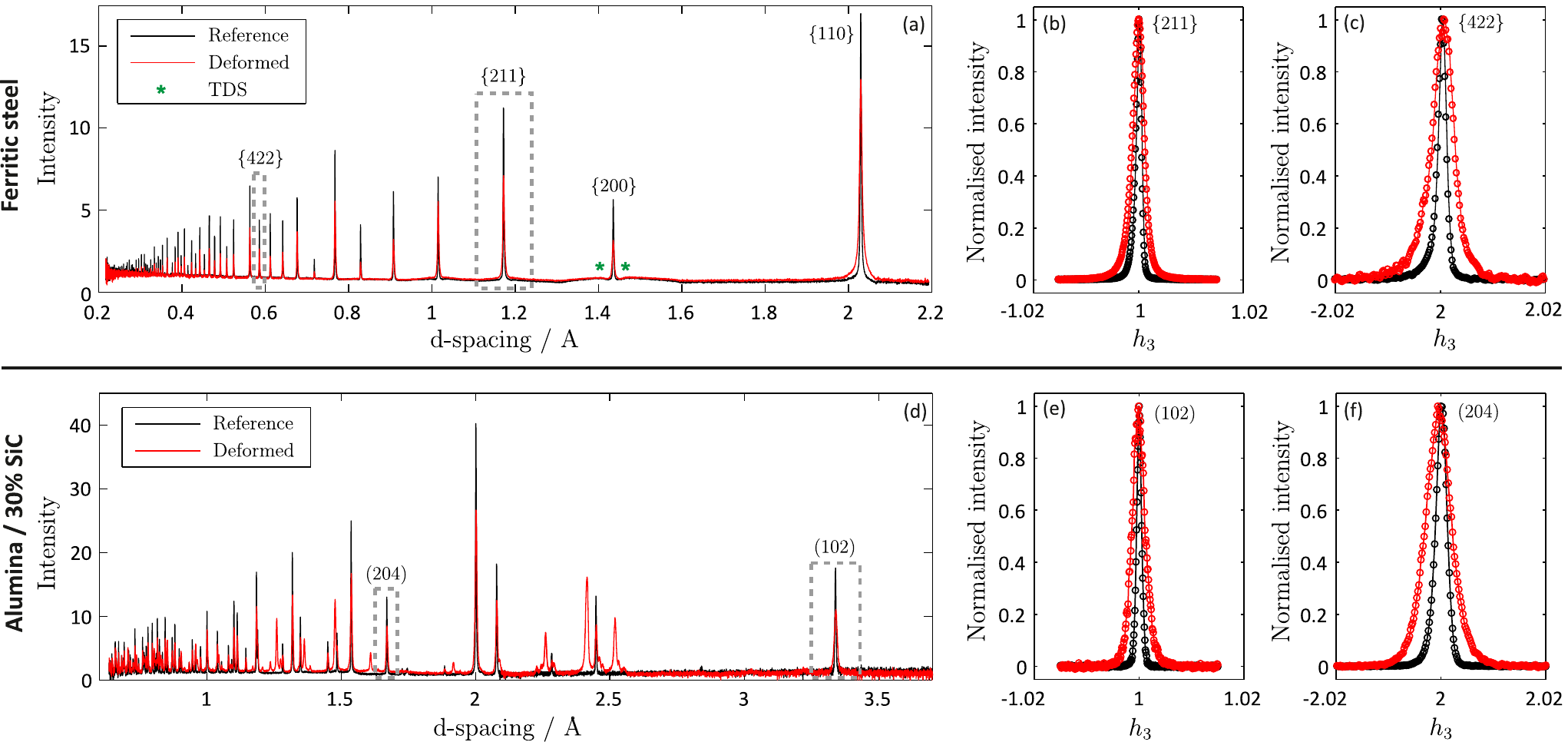}\\
    \caption{Raw data collected from HRPD instrument showing a ferritic steel and alumina-30\% SiC in the reference and deformed state, with (a) \& (d) full spectra, (b) \& (e) first order line profiles, and (c) \& (f) second order line profiles.}
    \label{Fig1}
\end{figure*}


\noindent where $A_n^S = N_n/N_3$ and $A_n^D = \langle \cos 2 \pi l Z_n \rangle$. If $n$ and $l$ are small, $Z_nl$ must also be small and the first two non zero terms from a Maclaurin series provide the following approximation
\begin{equation}
\langle \cos 2 \pi l Z_n \rangle   \rightarrow  1 - 2\pi^2l^2\langle Z_n^2 \rangle~~.
\end{equation}

\noindent In logarithm form, $\ln \langle \cos 2 \pi l Z_n \rangle = -  2\pi^2l^2\langle Z_n^2 \rangle$, so that Equation \ref{FourierProduct} becomes
\begin{equation}
\ln A_n(l) = \ln A_n^S -  2\pi^2l^2\langle Z_n^2 \rangle~~.
\end{equation}

\noindent As the domain size coefficient is independent of the order, $l$, and strain is dependent on $l$, the coefficients can be easily separated by plotting $\rm{ln} A_n(l)$ against $l^2$ with constant $n$ for 2 or more orders $l$ of the same reflection. The intercept gives the strain coefficient $A_n^S$ and the gradient gives $\langle Z_n^2 \rangle$. 

The $Z_n$ and $A_n^S$ values can be related to more conventional physical quantities as follows. The $Z_n$ values relate to pairs of unit cells separated by a distance $L=na_3$. The relative displacement of the unit cells due to the distortion, $\Delta L = a_3 Z_n$. The corresponding strain $\epsilon_L = \frac{\Delta L}{L} = \frac{Z_n}{n}$. It is therefore possible to use the $Z_n$ values to calculate the rms strain ($\epsilon^{1/2}$) normal to the diffracting planes and averaged over a distance $L$. As 
\begin{equation}
\label{rms_strain_eq}
\langle\varepsilon_L^2\rangle^\frac{1}{2} = \frac{\langle Z_n^2\rangle^\frac{1}{2}}{n}
\end{equation}

\noindent $L$ can be found by recognising that $a_3$ is the interplanar spacing for the relevant reflection. For the size coefficients Warren \cite{WarrenProgMetal} shows that
\begin{equation}
-\left(\frac{{\rm{d}} A_n^S}{{\rm{d}} n}\right)_{n=0} = \frac{1}{N_3}~~.
\end{equation}

\noindent Having obtained $N_3$ from the initial gradient of the graph of $A_n^S$ versus $n$, the mean crystallite size normal to the diffracting planes can therefore be calculated as $N_3a_3$.

\section{Method}

\subsection{Materials Studied}

Two materials have been studied; (1) a low carbon ferritic steel and (2) an alumina-30vol\% SiC composite. Their details are as follows:

A single phase ferritic steel `DX56'  was studied with nominal composition Fe-0.12C-0.5Si-0.6Mn-0.1P-0.045S-0.3Ti. The material had an initially weak crystallographic texture inherited from prior rolling during its processing and a mean grain size of 15\,$\rm{\micro}$m. The material was studied in an annealed, strain-free reference condition and in a deformed state. The latter was deformed via cold rolling parallel to the initial rolling direction to an engineering strain in the rolling direction of 0.5.

An alumina-30 vol\% SiC composite was made by ball milling Sumitomo AES-11C $\alpha$-alumina and F800 $\alpha$-SiC powders (d$_{50}$ = 6.5 $\micro$m) together in water using Mg-PSZ media with the aid of a dispersant (Allied Colloids Dispex A40) and spray dried. The powders were then hot pressed in a graphite die under Argon at 1700\,$^\circ$C to form a disc 30 mm diameter $\times$ 4\,mm thickness. The surface layers on both sides of the disc were ground off using diamond abrasives and the resulting disc had a relative density of $>$99\%. Microstructural examination showed similar features to the composites reported previously  \cite{Todd2004:1621}, in which the approximately equiaxed and angular SiC particles were essentially unchanged by incorporation in the composite. An alumina reference specimen was produced using the same methods but without the SiC addition.

\subsection{Time-of-flight diffraction}

All measurements were made using a 15\,mm $\times$ 15\,mm neutron beam. The neutrons sampled by the backscattered detectors measured direct strains at angles between 2$^\circ$ and 10$^\circ$ to the incoming beam. The steel specimens each had a cubic geometry of 8 $\times$ 8 $\times$ 8\,mm$^3$. As the samples were prepared from thin sheets, this volume was achieved by layering successive coupons and affixing them with cyanoacrylate adhesive. Samples were placed with the incoming beam direction parallel to the rolling direction of the sample. This sample orientation relative to the backscatter detectors provided diffraction data from planes with normals approximately parallel to the rolling direction, providing results that can be related to this deformation axis. The alumina-30\% SiC sample was positioned with its surface plane normal at 45$^\circ$ to the incoming beam. The specimen was positioned such that all parts of the beam passed through it. The inclination of the specimen to the beam ensured the measurement of strain values between the extremes of the small amount of anisotropy caused by the alignment of slightly elongated SiC particles during hot pressing \cite{Todd2004:1621}. 

From the measured diffraction data, several preparatory steps were applied to obtain line profile information suitable for the Warren-Averbach analysis. Time-of-flight/$d$-spacings from the raw data were firstly converted to $h_3$ space (Eq. 9). Next, individual line profiles were extracted from the diffraction spectra. These had a $h_3$ range of $l - \Delta h_3$ to $l + \Delta h_3$ where $l$ is the order of refection studied (i.e. 1, 2, ...) and $\Delta h_3$ is the interval. The size of these intervals corresponded to $4$ times the full width at half maximum (FWHM) as recommended by Schwartz and Cohen \cite{Cohen77}. 

Prior to the deconvolution procedure, some practical steps were taken: (1) a uniform $h_3$ data point spacing was set; a constant spacing was determined and corresponded to interpolated intensities, here using a 1-dimensional linear interpolation scheme, (2) any background intensity was removed; a linear function was fitted here to a number of data points at the extremes of the range ($\sim$$0.05\times\Delta h_3$), then deducted from the line profile intensity, and (3) the background corrected intensity data were normalised. 

Each experimentally measured diffraction line profile, $h(h_3)$ is described by the convolution of the specimen line profile, $f(h_3)$, and instrumental line profile, $g(h_3)$. This is written as
\begin{equation}
\label{conv}
h(h_3) = f(h_3)\otimes g(h_3)
\end{equation}

\noindent where $\otimes$ is the convolution operator. As it is essential to interrogate information only from the sample, $f(h_3)$ must be separated. Equation \ref{conv} may be solved for $f(h_3)$ using the convolution theorem. Using the Stokes method \cite{Stokes}, the Fourier coefficients of $f(h_3)$ are given by
\begin{equation}
\label{realFourier}
A_n = \frac{H_c(n)G_c(n)+H_s(n)G_s(n)}{G_c^2(n)+G_s^2(n)}
\end{equation}
\begin{equation}
B_n = \frac{H_s(n)G_c(n) - H_c(n)G_s(n)}{G_c^2(n)+G_s^2(n)}
\end{equation}

\noindent where $H(n)$ and $G(n)$ are the Fourier coefficients of the uncorrected sample peak and the instrumental line profiles, respectively, for a given harmonic number, $n$, and subscripts $c$ and $s$ refer to cosine and sine components, respectively. In principle, the true sample profile $f(h_3) = I(h_3)$ can be reconstructed using Equation 10, though this is not necessary for the Warren-Averbach analysis. 

In the studies here, $h(h_3)$ is a deformed sample, (the cold-rolled ferritic steel or the alumina-30 vol\% SiC composite) and $g(h_3)$ is an undeformed sample (annealed ferritic steel or alumina reference). The Fourier components were calculated by:
\begin{equation}
\label{Four1}
H_c(n) = \int_ {l - \Delta h_3}^{l + \Delta h_3}h(h_3) \cos (2\pi n h_3)\,{\rm{d}} h_3
\end{equation}
\begin{equation}
\label{Four2}
H_s(n) = \int_ {l - \Delta h_3}^{l + \Delta h_3}h(h_3) \sin (2\pi n h_3)\, {\rm{d}} h_3
\end{equation}
\begin{equation}
\label{Four3}
G_c(n) = \int_ {l - \Delta h_3}^{l + \Delta h_3}g(h_3) \cos (2\pi n h_3)\, {\rm{d}} h_3
\end{equation}
\begin{equation}
\label{Four4}
G_s(n) = \int_ {l - \Delta h_3}^{l + \Delta h_3}g(h_3) \sin (2\pi n h_3)\, {\rm{d}} h_3 ~~.
\end{equation}

\noindent These components were normalised with respect to $H^0$ and $G^0$ which were obtained by setting $n=0$ in Equations \ref{Four1} and \ref{Four3}, respectively.  All integrations were performed using Simpson's rule. These coefficients were calculated only for values of $n$ for which the approximation of ${\rm{ln}} \langle \cos 2 \pi l Z_n \rangle$ in Equation 15 led to an error of less than 10\% in the second order peak. 

\subsection{EBSD-based dislocation density validation}

The ferritic steel in the reference and deformed states was characterised using EBSD to obtain spatially resolved crystallographic orientations. Data were obtained using a JEOL6500F (FEG) scanning electron microscope operating at an accelerating voltage of 20\,kV with a probe current of $\sim$15\,nA. Diffraction patterns were acquired with a TSL Digiview II camera at a 1000$\times$1000 pixel$^2$ resolution at a $\sim$1\,s acquisition time.

Geometrically necessary dislocation (GND) density estimates can be obtained via measurement of EBSD pattern shifts \cite{Wilkinson2010} to obtain lattice curvatures that can be subsequently solved through analysis of the Nye tensor \cite{Nye}. The process of acquiring an EBSD map will typically obtain information from a finite patch of microstructure. If multiple grains lie within this finite patch, all crystals, irrespective of orientation will be identified. This is not equivalent to information obtained from time of flight diffraction; only a subset of an EBSD-analysed grain set will have orientations that would obey Bragg's law. A simple method is described here to extract the EBSD estimates of dislocation densities from the orientations equivalent to those measured from time of flight diffraction. 

The crystal orientation from the EBSD measurements is described through the rotation matrix $\bf{G} = \bf{R}_x(\phi_1)R_z(\Phi)R_x(\phi_2)$ where $\phi_1$, $\Phi$, $\phi_2$ are the Euler angular operations. The plane normal to each reflection ($hkl$) observed from neutron diffraction, from the crystal reference frame, can be described in the global reference frame via
\begin{equation}
	\bf{r} = \bf{G}
	 \begin{bmatrix}
	h \\ k \\ l
	\end{bmatrix}.
\end{equation}

The HRPD backscatter detector comprises eight octets that are positioned radially about the incident neutron beam. Each octet is out of plane, accepting a diffraction angle of $160^\circ < 2\theta <176^\circ$. Assuming the incident beam is in the $z$ direction, the radial positions have coordinates $x,y$. From the vector, $\bf{r}$, this corresponds to components $r_x$ and $r_y$. For any diffraction vector, $\bf{Q}$, detected, the component in the beam direction is given by
\begin{equation}
Q_z = \frac{\sqrt{r^2_x + r^2_y}}{\sin2\theta}~~.
\end{equation}

\noindent The diffraction vector, $\bf{Q}$, is given by
\begin{equation}
\bf{Q} = 
	 \begin{bmatrix}
	r_x \\ r_y \\ Q_z
	\end{bmatrix}.
\end{equation}

\noindent If the vector $\bf{r}$ is within the range of vectors allowed for $\bf{Q}$ (for the allowed $2\theta$ range), the crystal will diffract. Practically, this enables each data point within an EBSD map to be analysed, to deduce whether this location has an orientation that would permit time-of-flight diffraction, for a given reflection ($hkl$). 

\section{Results}

The experimental raw data are shown in Fig.\,\ref{Fig1}a for the ferritic steel and Fig.\,\ref{Fig1}d for the alumina-30\% SiC. Traces are shown in the reference state (black) and with deformation (red). The first and second order reflections used for line broadening analysis are also given, comprising the \{211\} \& \{422\} (Fig.\,\ref{Fig1}b \& Fig.\,\ref{Fig1}c) for the steel and \{102\} \& \{204\} (Fig.\,\ref{Fig1}e \& Fig.\,\ref{Fig1}f) for the alumina composite. The normalised intensities of these individual reflections have been plotted with respect to the dimensionless parameter, $h_3$. 

The corrected Fourier coefficients, $A_n$  for the sample were calculated using Equation 19. This is shown in Fig. 3a \& e for the steel and ceramic samples, respectively. For each value of $n$, a plot of $\ln A_n(l)$ versus $l^2$ yields $\ln A_n^\mathrm{S}$ at the intercept and the gradient is $-2\pi^2\langle Z_n^2 \rangle$, as shown in Equation 15. These are shown in Fig. 3b \& f. To deduce the crystallite size from the line broadening, $A_n^\mathrm{S}$ is plotted against $L$, enabling the mean column length, $N_3$, to be obtained at the intercept of the initial slope on the axis of abscissa. This is shown in Fig. 3c \& g for the steel and composite samples, respectively. In both cases the $A_n^S$ vs. $L$ plots are essentially horizontal, with all values close to 1. This shows that the crystallite sizes were so large that their effect did not contribute significantly to the peak breadth. Meaningful crystallite sizes cannot therefore be extracted. This was confirmed by reconstructing the deconvoluted sample peaks, using Equation 10. When plotted as a function of $h_3/l$, the peaks from the first and second order peaks were coincident.


\begin{figure*}
     \centering
     \includegraphics[width=\textwidth]{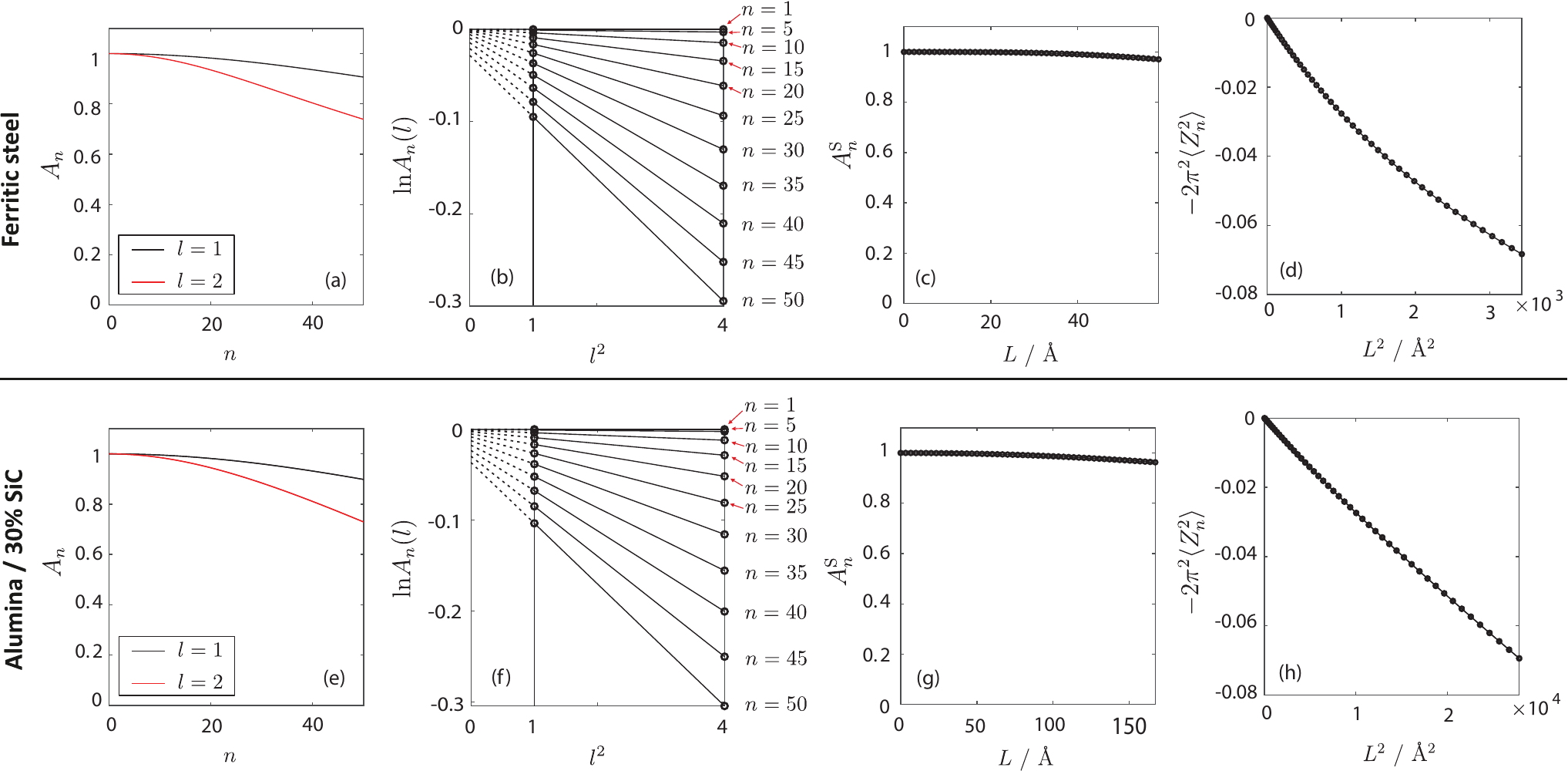}\\
     \caption{Calculated Fourier coefficients for the deconvoluted $f(h_3)$ line profiles for the first order ($l=1$) and second order ($l=2$) reflections, for ferritic steel (a) and alumina-30\% SiC (b). Particle size and strain effects can be separated with the logarithmic plots shown in (b) and (f) for multiple orders. Data used for particle size calculation are shown in (c) \& (g) and strain calculations can be calculated from the relationship between $-2\pi^2\langle Z_n^2\rangle$ and $L^2$, (d) \& (h).}
     \label{Fig3}
\end{figure*}


Using the calculated $-2\pi^2\langle Z_n^2 \rangle$ terms, the strain contribution to broadening was calculated. The relationship between this term and distance squared ($L^2$) is shown in Fig.\,\ref{Fig3}d \& (h) for the steel and ceramic samples, respectively. The steel sample shows a distinct concave $-2\pi^2\langle Z_n^2 \rangle$ - $L^2$ trend whereas the ceramic has a near perfect proportional $-2\pi^2\langle Z_n^2 \rangle$ - $L^2$ relationship. 

\begin{figure}
     \centering
     \includegraphics[width=60mm]{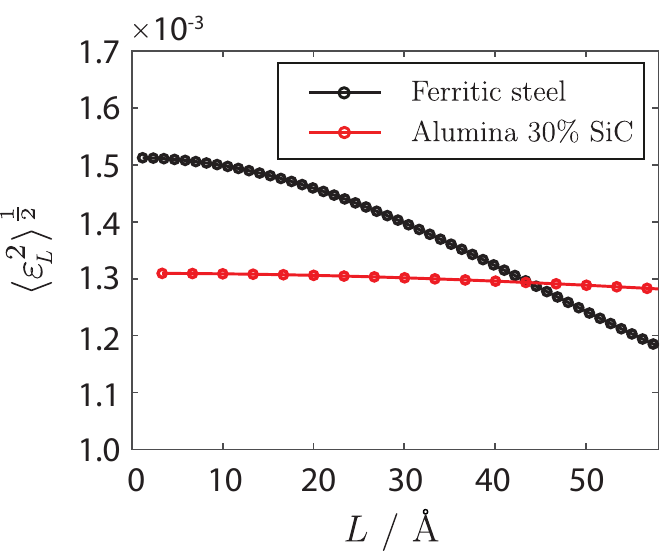}\\
     \caption{Root mean squared strain for ferritic steel and alumina composite samples.}
     \label{Fig4}
\end{figure}

Using Equation \ref{rms_strain_eq}, the rms (root mean squared) strain, $\langle \varepsilon_L^2 \rangle ^\frac{1}{2}$, was next calculated with respect to averaging distance, $L$. The results are shown in Fig.\,\ref{Fig4}. The magnitude of $\langle \varepsilon_L^2 \rangle ^\frac{1}{2}$ for the ferritic steel is shown to decrease as a function of increasing averaging distance indicating that the strains caused by the dislocations vary significantly over the range of lengths, $L$ sampled. This will be used in Section 5.1 to estimate the dislocation density.

The rms strain in the alumina-SiC composite remains approximately constant at $\sim1.3\times10^{-3}$ over the full range of averaging distance, $L$. This is because the thermal residual strains vary on the scale of the microstructure i.e. several micrometres, which is much greater than the length scales of $\sim10$\,nm accessible to the Warren-Averbach analysis. The large grain size is also the reason that the particle size broadening is negligible for this specimen.

\section{Discussion}

\subsection{Analysis and validation of results: steel}

A measurement of rms strain with respect to increasing averaging distance can be used to estimate the dislocation density \cite{Wilkens:1970, Groma:gk0105}; this has been performed for the steel sample examined in this study. A dislocation density estimate that assumes straight, parallel, randomly distributed screw dislocations was proposed by Krivoglaz and Ryaboshapk \cite{Krivoglaz1963}:
\begin{equation}
\label{DisDens}
	\langle\varepsilon_L^2\rangle \cong \frac{\rho C b^2}{4 \pi} \ln \left(\frac{R_e}{L}\right)
\end{equation}

\noindent where $\rho$ is the dislocation density, $C$ is a contrast factor, $b$ is the Burgers vector and $R_\mathrm{e}$ is the dislocation cutoff radius. The expression assumes $R_\mathrm{e}>L$.  The contrast factor is dependent on the diffraction vector, dislocation line vector and Burgers vector and may be calculated numerically \cite{Wilkens:1970}. For a cubic system, an average contrast factor, $\bar{C}$ can be used in a simpler formulation \cite{Ungar:1999}; this method accounts for the elastic anisotropy of the material and the diffraction vector relative to the slip systems, assumed to be $\{110\}\langle\bar{1}11\rangle$ slip alone for the ferritic steel. For steel, the average contrast factor calculated for the \{211\} reflection was 0.1040 assuming screw dislocations only. As the ferritic steel studied has a BCC crystal structure, deformation is typically mediated by screw dislocations due to the availability of multiple slips planes (\{110\}, \{112\} or \{123\}) onto which the characteristic non-planar cores dislocations cores can exist whilst preserving the slip direction, $\langle111\rangle$ \cite{HullBacon}. Therefore, assuming all dislocations are of screw type is deemed appropriate as an estimate. This may be extended to account for edge/screw dislocation ratio \cite{ungar96}. Further details of the Equation \ref{DisDens} approximation including a comprehensive description of the relevant supporting literature is available elsewhere (e.g. \cite{Ungar2001}).

\begin{figure}[h!]
     \centering
     \includegraphics[width=60mm]{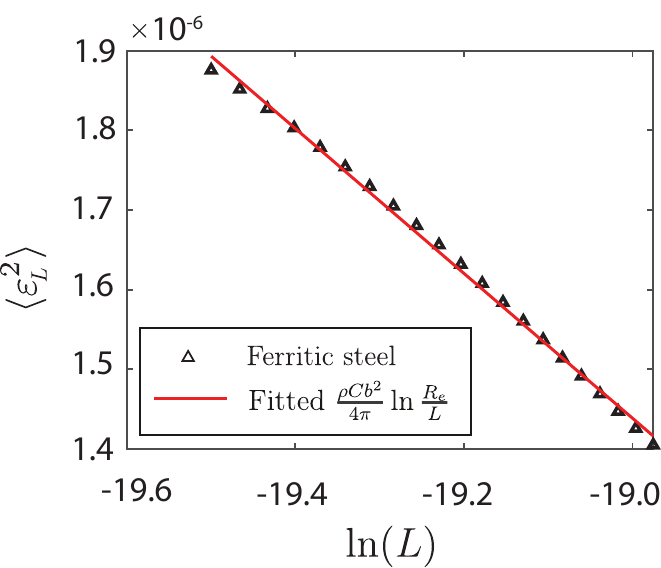}\\
     \caption{Fitted ferritic steel data for dislocation density calculation.}
     \label{Fig5}
\end{figure}

If one plots $\langle\varepsilon_L^2\rangle$ versus $\ln L$ the gradient is $-\frac{\rho C b^2}{4\pi}$ and $R_\mathrm{e}$ can be calculated from the intercept. An example fit for the ferritic steel is shown in Fig.\,\ref{Fig5}. The measured dislocation density is $1.78\times10^{15} \pm 0.05\times10^{15}$ \,m$^{-2}$ and $R_\mathrm{e}$ is $27 \pm 18$\,nm (errors calculated from 95\% confidence of fit). 

Williamson and Smallman \cite{Smallman1956} presented an alternative dislocation density estimation with no $L$ dependency; 
\begin{equation}
\rho = \frac{k}{F} \frac{\xi^2}{b^2}
\end{equation}

\noindent where $k$ is a crystal structure dependent constant; this is 14.4 for the BCC crystal structure \cite{Smallman1956} and $F$ is an interaction factor of order 1 for widely distributed dislocations. This value assumes an idealised organisation of dislocations; the metal comprises a set of blocks with a single dislocation located at each block boundary. $\xi$ is the strain distribution integral breadth. Assuming a Gaussian strain distribution this corresponds to $\sqrt{2\pi}\langle \varepsilon_L^2 \rangle^\frac{1}{2} $ when $n=1$. This method gives a dislocation density estimate in the ferritic steel of $3.3\times10^{15}$\,m$^{-2}$.

\begin{figure*}
     \centering
     \includegraphics[width=170mm]{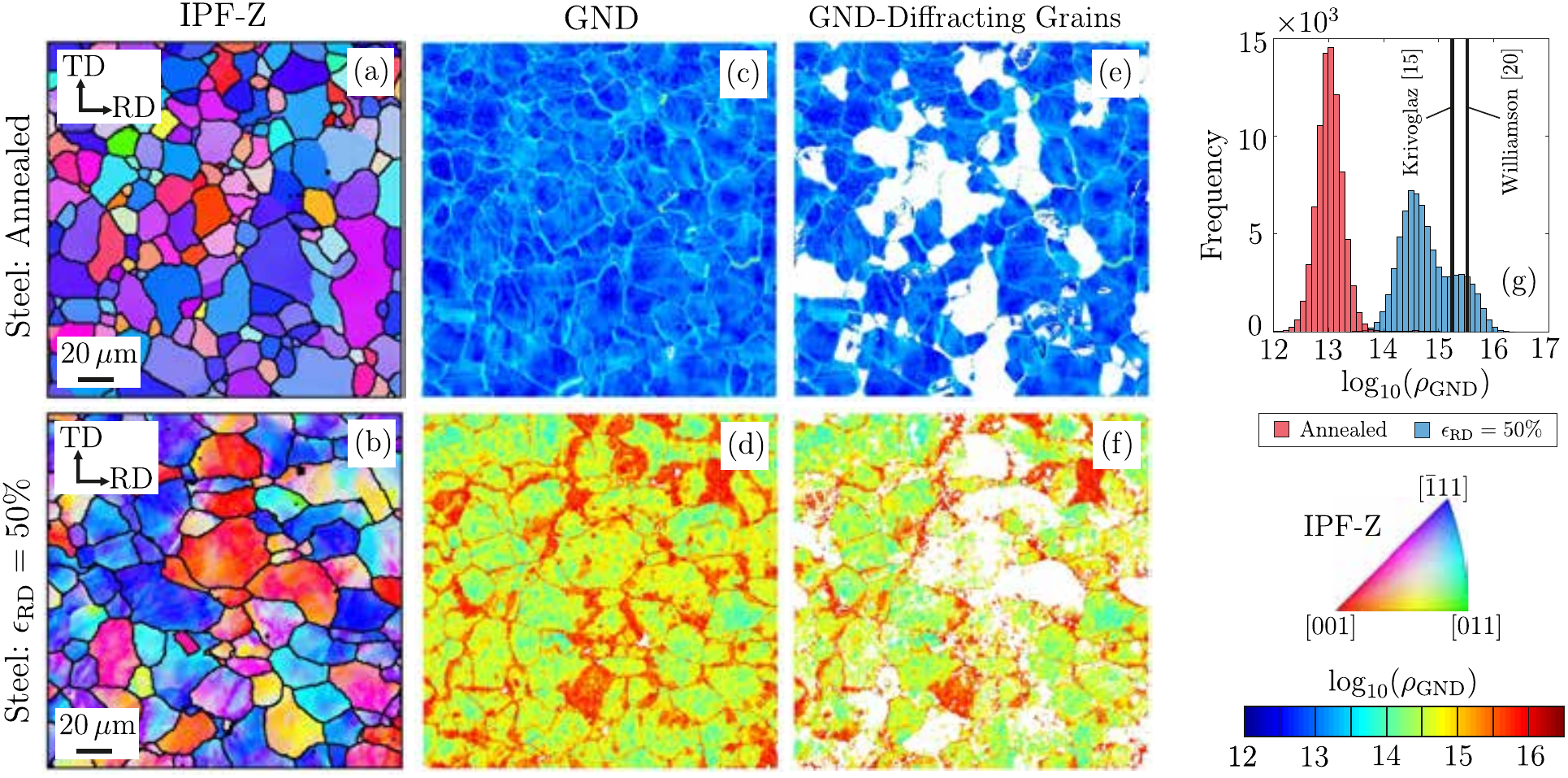}\\
     \caption{EBSD IPF-Z plots for ferritic steel in the (a) annealed and (b) cold-rolled states. The corresponding GND density maps are shown in (c) \& (d). In (e) \& (f), the GND density maps are plotted only for regions that satisfy the HRPD geometry for diffraction from the \{211\} lattice planes. GND density histograms of the annealed and cold-rolled steels are also shown (g) together with peak-broadening estimates of dislocation density.}
     \label{Fig6}
\end{figure*}


A validation test is given here for the dislocation density measurement of the steel sample. The undeformed and deformed states of the ferritic steel examined with time of flight neutrons were characterised with EBSD. Inverse pole figure plots with respect to the $Z$-direction (normal direction, ND) are shown in Fig.\,\ref{Fig6}a \& Fig.\,\ref{Fig6}b. Grain boundaries of misorientation $>$10$^\circ$ are shown with continuous black lines. The corresponding GND density maps are shown in Fig.\,\ref{Fig6}c \& Fig.\,\ref{Fig6}d. The undeformed sample possesses GND densities in the 10$^{12}$ - 10$^{13}$\,m$^{-2}$ region, which is several orders of magnitude smaller than the deformed specimen (10$^{14}$ - 10$^{16}$\,m$^{-2}$). The regions of these maps that have crystal orientations that would diffract from the \{211\} lattice planes using the method proposed earlier are shown in Fig.\,\ref{Fig6}e \& Fig.\,\ref{Fig6}f. To aid a comparison of this data to dislocation density estimates made via line profile broadening, histograms of the GND densities (for the subset of grains that satisfy the HRPD \{211\} reflection diffraction condition), for each of the deformation states, are shown in Fig. \ref{Fig6}g. The dislocation density estimates obtained from time of flight line broadening are superimposed. These values sits within the distribution of the deformed specimen, and are both slightly higher than the mean GND density of $1.2 \times10^{15}$\,m$^{-2}$. A higher value is expected from Warren-Averbach analysis because this will sample statistically stored as well as geometrically necessary dislocations. The neutron results are therefore good estimators for the mean dislocation density of plastically deformed metals. Data presented in this manner emphasise that a single value for the dislocation density may be a good estimator of the mean dislocation density, but neglects the dislocation density range, or indeed dislocation density hotspots which may be of greater interest in a structural metallic material.

Although the dislocation density estimates from time of flight diffraction show good agreement to the GND results, the GND analysis from EBSD data shows that the ferritic steel can exhibit highly inhomogeneous deformation (Fig 6d), with certain grains accumulating much higher dislocation densities than neighbours. This was evident from the the bimodal distribution of the estimated GND density obtained (shown in Fig.\,\ref{Fig6}g). Both models relating rms strain to dislocation density assume a uniform dislocation distribution and such results should therefore be treated as approximate.

\subsection{Analysis and validation of results: alumina-SiC composite}

In order to test the validity of the strain analysis for the alumina-SiC composite, we estimate an approximate relationship between the mean strain in the alumina matrix of the composite, which can be measured straightforwardly from diffraction peak shifts relative to the pure alumina reference specimen, and the rms strain measured using the Warren-Averbach analysis. Using the mean field approximation of Tanaka and Mori \cite{TANAKA1970931}, the thermal residual stress in the vicinity of a particle in a composite consisting of a finite volume fraction $f$ of particles dispersed in a matrix can be approximated by the stress which would result from the presence of a single particle in an infinite matrix, plus a background stress which arises due to the actual, finite volume fraction of particles, and which acts as though applied externally to maintain the force balance within the composite. Assuming the particles to be spherical, the stress $\sigma_{p\infty}$ within a single particle in an infinite matrix is hydrostatic. A simplification can be introduced by noticing that the bulk moduli of alumina and SiC are the same to within a few \% (248\,GPa and 244\,GPa using data in ref \cite{RIT1997}). The background stress is therefore uniform and hydrostatic with magnitude $-f\sigma_{p\infty}$.  This gives principal stresses in the matrix at its interface with the particle \cite{Timoshenko}:
\begin{equation}
\label{RIT_Strain1}
\sigma_{mr{\rm (int)}} = \sigma_{p\infty}(1-f)
\end{equation}
\begin{equation}
\label{RIT_Strain2}
\sigma_{m\theta{\rm (int)}} = \sigma_{m\phi {\rm (int)}} = \sigma_{p\infty}\left(-\frac{1}{2}-f\right)
\end{equation}

\noindent where $r$, $\theta$, $\phi$ are spherical polar coordinates. The difference between the corresponding principal strains gives an estimate of the maximum range of normal strains sampled during neutron diffraction:
\begin{equation}
\label{RIT_Strain3}
\Delta\varepsilon_{\rm max} = -\frac{3\sigma_{p\infty}(1-\nu_m)}{2E_m}
\end{equation}

\noindent where $E_m$ and $\nu_m$ are the Young's modulus and Poisson's ratio of the matrix.  The mean strain in the matrix is:
\begin{equation}
\label{RIT_Strain4}
\langle \bar{\varepsilon}_m \rangle = -\frac{f\sigma_{p\infty}(1-2\nu_m)}{E_m}
\end{equation}

\noindent and combining Equations \ref{RIT_Strain3} and \ref{RIT_Strain4} gives an estimate of the maximum half width of the strain distribution around a central position:
\begin{equation}
\label{RIT_Strain5}
\pm \frac{\Delta\varepsilon_{\rm max}}{2} = \frac{3 \langle \bar{\varepsilon}_m \rangle (1+\nu_m)}{4f (1-2\nu_m) }~~.
\end{equation}

\noindent $\langle \bar{\varepsilon}_m \rangle$ was measured from the diffraction peak shift between the composite specimen and the alumina reference, and was equal to $3.4 \pm 0.5\times10^{-4}$. With $f = 0.3$ and $\nu_m = 0.23$, equation 5 gives $\Delta\varepsilon_{\rm max}/2 = 1.9\times10^{-3}$. This maximum half width is expected to be of the same order, but a little larger than the rms strain $\langle \varepsilon_L^2 \rangle^\frac{1}{2} $.  The value obtained is about 50\% larger than the rms strain, $\langle \varepsilon_L^2 \rangle^\frac{1}{2} $ of $1.3 \times10^{-3}$ measured using the Warren-Averbach analysis, giving strong support to the validity of the method.

\subsection{Experimental considerations}

The Warren theory applies in principle to any set of planes. However, not all reflections were found to be suitable for analysis in this study. For the ferritic steel, selecting the \{110\} reflection was unsuitable as the background was highly non-uniform. Accounting for this with confidence was deemed to be too difficult. Evidence of thermal diffuse scattering (TDS) in the vicinity of low index reflections was apparent. In particular these scattering artefacts were seen to flank the shoulders of the \{200\} reflection in the ferritic steel, labelled ($\star$) in Fig.\ref{Fig1}a. Whilst the intensity of such scattering increases with smaller $d$-spacing \cite{IntroResidualStress}, for a polycrystalline sample this incoherent scattering becomes less structured, becoming more like a uniform background, as the volume of reciprocal space over which integration occurs increases. For the \{211\} \& \{422\} reflections used, there was no clear evidence of TDS in the vicinity of the line profiles. Considering higher index reflections is valid, but their use becomes increasingly difficult as (1) the line profiles do not decay to the background level before encroaching on a neighbouring reflection in close proximity, and (2) noisy (low signal to noise ratio) line profiles provide unreliable values of the Fourier coefficients, $A_n$, making subsequent analysis less accurate. The methodology adopted in this study, does, however include strategies to judiciously select appropriate reflections to obtain valid results.

For the \{211\} type reflections chosen in this study the Schwartz and Cohen \cite{Cohen77} recommendation of $\sim4$ FWHM for the tails could be achieved without encroaching on the neighbouring peaks and the above validation indicates that meaningful results were obtained. However, the $\langle\varepsilon_L^2\rangle^\frac{1}{2}$ estimates for low $n$ showed sensitivity to variations in the $h$ range used in the analysis, indicating the validity can only just be achieved within the conflicting requirements described above. For a range of more than FWHMs it was evident that tails of neighbouring peaks were being included in the analysis. Variations in the low $n$ strain values were also found for tail widths of less than 4 FWHM, this indicates that the tails were being truncated. This indicates that low $L$ values of rms strain are less reliable and the dislocation analysis, using Equation \ref{DisDens}, cannot be reliably obtained if the corresponding rms values are included. Rothman and Cohen \cite{Rothman} found that strains around dislocations must yield a linear decrease in mean-square strains, $\langle\varepsilon_L^2\rangle$, with averaging distance, $L$. In practice, this was not observed for the steel sample for low $L$ values, owing to the longs tails, and thus the dislocation density approximation from Equation \ref{DisDens}, with results shown in Fig. \ref{Fig6}, used values only where a linear $\langle\varepsilon_L^2\rangle$ - $L$ relationship exists. 

\section{Conclusions}  

The line profile broadening measured from high resolution time-of-flight neutron diffraction data has been quantified from deformed samples of a single phase ferritic steel and an alumina-30\% SiC ceramic composite. The following conclusions can be drawn:

\begin{enumerate}

\item The Warren-Averbach theory for analysing peak broadening from X-ray diffraction data has been shown to be suitable for time of flight data.

\item The data analysis method was capable of describing the root mean square strain, $\langle\varepsilon_L^2\rangle^\frac{1}{2}$, of two materials deforming in different ways. The plastically deformed steel exhibits a rapid decay of strain with averaging distance, characteristic of plastic deformation. A composite ceramic exhibited an Al$_2$O$_3$ phase subjected to thermal residual strains varying on the scale of the microstructure, gave a near uniform $\langle\varepsilon_L^2\rangle^\frac{1}{2}$ with distance over the measured range. 

\item Obtaining an appropriate estimate of the dislocation density and rms strain requires a best practice methodology of selecting an appropriate $h_3$ interval that is both sufficiently wide to capture the line profile shape, but not too wide that neighbouring reflections or other background artefacts are included. The analysis should also consider the sensitivity of low $L$ rms strain values due to the tails of the broadened line profile.

\item A method to compare the dislocation density estimated from the Warren-Averbach method and lattice curvatures measured from EBSD characterisation has been devised that accounts for grains in each case that are similarly orientated. Good agreement was seen between the dislocation density estimated between the two methods for deformed ferritic steel. 

\item A numerical estimate of the uniform elastic strain in an Al$_2$O$_3$ - SiC composite provides good agreement with the rms strain measured by the time of flight Warren-Averbach analysis method.

\end{enumerate}

\section{Data Statement}
An open-source Matlab version of the time of flight Warren-Averbach analysis code is freely available for download via GitHub ({https://github.com/d-m-collins/ToF-WarrenAverbach}). The ferritic steel raw data presented in this paper is made available as an example.

\section{Acknowledgments}

Financial support of this work was provided by the EPSRC (grant EP/I021043/1) and steel provided by BMW-MINI. Beamline time at ISIS (RB1220050) and experimental support from Dr S. Kabra \& Dr A. Daoud-Aladine is gratefully acknowledged. With thanks also to Prof. Andrew Goodwin for his guidance in TDS observations.


\section*{References}

\bibliographystyle{model3-num-names}
\bibliography{WarrenTheory}

\end{document}